\documentclass[lettersize,journal]{IEEEtran}
\usepackage{amsmath,amsfonts}
\usepackage{algorithmic}
\usepackage{algorithm}
\usepackage{array}
\usepackage[caption=false,font=normalsize,labelfont=sf,textfont=sf]{subfig}
\usepackage{textcomp}
\usepackage{stfloats}
\usepackage{url}
\usepackage{verbatim}
\usepackage{graphicx}
\usepackage{cite}
\usepackage{dsfont}
\usepackage{multirow}
\usepackage{booktabs}
\usepackage{multirow}
\usepackage{makecell}
\begin{document}

\title{SEI-SHIELD: Robust Specific Emitter Identification Under Label Noise Via Self-Supervised Filtering and Iterative Rescue}

\author{Ruixiang Zhang, Zinan Zhou, Yezhuo Zhang, Guangyu Li and Xuanpeng Li
\thanks{Copyright may be transferred without notice, after which this version may
no longer be accessible. \textit{Corresponding author: Xuanpeng Li}}
\thanks{Ruixiang Zhang, Zinan Zhou, Yezhuo Zhang and Xuanpeng Li are with the School of Instrument Science and Engineering, Southeast University, Nanjing, 210096, Jiangsu, China (e-mail: \protect\url{zhang_ruixiang@seu.edu.cn}; \protect\url{zhouzinan919@seu.edu.cn}; \protect\url{zhang_yezhuo@seu.edu.cn}; \protect\url{li_xuanpeng@seu.edu.cn}).}
\thanks{Guangyu Li is with the School of Computer Science and Engineering, University of Science and Technology, Nanjing, 210094, Jiangsu, China (email: guangyu.li2017@njust.edu.cn).}
}

\markboth{Journal of \LaTeX\ Class Files,~Vol.~14, No.~8, August~2021}%
{Shell \MakeLowercase{\textit{et al.}}: A Sample Article Using IEEEtran.cls for IEEE Journals}

\IEEEpubid{0000--0000/00\$00.00~\copyright~2021 IEEE}
\maketitle

\begin{abstract}
Specific Emitter Identification (SEI) provides physical-layer device authentication for wireless communications and Internet of Things (IoT) systems.
While deep learning (DL) has significantly advanced SEI performance, label noise severely degrades system reliability in non-cooperative environments.
Label noise originates from channel-induced ambiguities, annotation errors, and deliberate data poisoning by intelligent jammers injecting misleading signals.
While recent SEI methods attempt to mitigate label noise, they fundamentally rely on corrupted supervised signals to guide sample selection, inevitably leading to confirmation bias and suboptimal feature spaces.
To address this challenge, we propose SEI-SHIELD, a robust SEI framework that integrates self-supervised contrastive pre-training with iterative sample selection. 
Specifically, SEI-SHIELD employs Momentum Contrast (MoCo) with RF-tailored augmentations to extract intrinsically robust, label-independent representations directly from complex-valued I/Q signals.
In addition, K-nearest neighbors (KNN)-based noise filtering identifies corrupted samples through neighborhood label consistency analysis in the learned feature space.
Furthermore, an iterative rescue mechanism using prediction confidence and prototype cosine similarity progressively recovers correctly labeled hard samples inadvertently discarded during filtering.
Comprehensive experiments on the POWDER and ORACLE datasets demonstrate that SEI-SHIELD achieves state-of-the-art (SOTA) accuracy under various noise rates, substantially outperforming existing noise-robust paradigms, including advanced regularization techniques and sample selection frameworks.
\end{abstract}


\section{Introduction}
\IEEEPARstart{W}{ith} the rapid development of wireless communication technologies and the proliferation of Internet of Things (IoT) devices, electromagnetic spectrum security has become increasingly prominent~\cite{10546997, 11024041}.
In this context, Specific Emitter Identification (SEI)~\cite{MAMC, 10596321} based on Radio Frequency Fingerprints (RFFs) has emerged as a critical physical-layer authentication method for enhancing communication security.
SEI exploits the unique hardware imperfections inherent in wireless transmitters, such as I/Q imbalance, phase noise, and power amplifier nonlinearity~\cite{9069935}, to uniquely identify individual devices without relying on cryptographic protocols.
This capability provides a complementary security layer for spectrum monitoring, access control, and defense against adversarial threats.

While deep learning has substantially advanced SEI performance, existing methods~\cite{9448105, 10285131, 9846906} primarily focus on maximizing training accuracy under the assumption of clean labels, overlooking the prevalent label noise in practical deployments.
This oversight leads to severe security vulnerabilities: models trained on noisy labels tend to memorize noisy patterns rather than learning true RFFs~\cite{zhang2017understanding}, producing overconfident but incorrect predictions that undermine the trustworthiness of physical-layer authentication systems.
Moreover, these errors can propagate through the training pipeline, making the system highly susceptible to adversarial data poisoning, where intelligent jammers deliberately inject misleading signals to corrupt training databases~\cite{11052286, 10539332}.
Consequently, developing label noise-resilient training frameworks is no longer merely an algorithmic optimization, but a fundamental security imperative to preserve the integrity of SEI systems.

\begin{figure}[!t]
    \centering
    \includegraphics[width=\columnwidth]{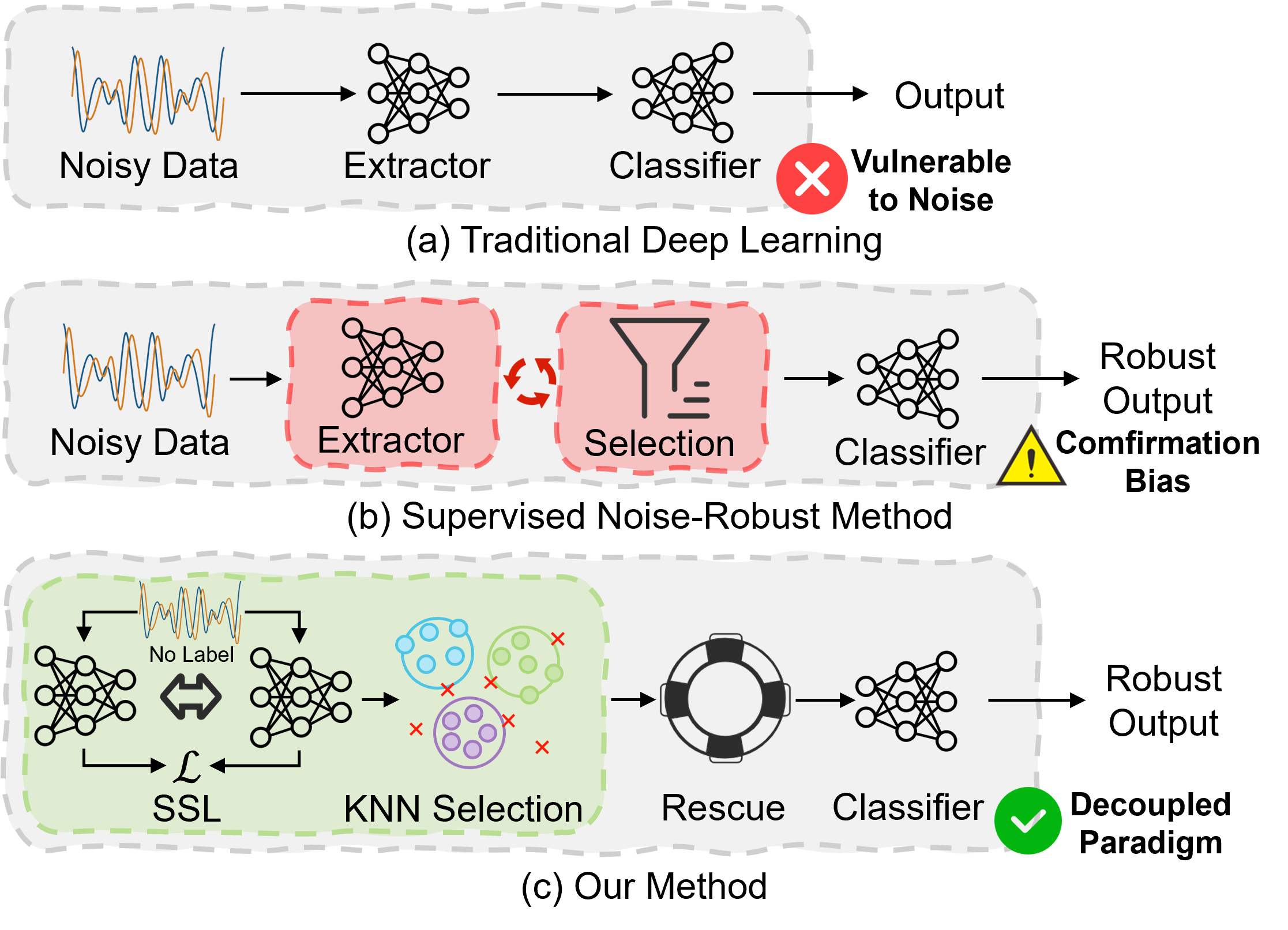}
    
    \caption{The paradigm evolution of label-noise robust SEI methods. (a) Traditional DL-based SEI. (b) Supervised Noise-Robust SEI. (c) SEI-SHIELD: Decoupled Paradigm.}
    \label{fig:paradigm_evolution}
\end{figure}

\IEEEpubidadjcol

Driven by this security imperative, a few pioneering works have attempted to mitigate label noise in SEI.
For instance, the Sample Selection and Regularization (SSR) method employs a two-stage adaptive selection strategy driven by confidence learning~\cite{SSR-SEI}. 
Similarly, the Noise-Robust SEI (NR-SEI) approach utilizes Jensen-Shannon divergence and clustering to partition data for semi-supervised training~\cite{NR-SEI}.
Despite their contributions, these methods exhibit critical limitations in highly corrupted environments.
First, they fundamentally rely on supervised training signals to guide sample selection—a paradoxical approach, as the supervision itself becomes unreliable at high noise rates~\cite{han2018co, arazo2019unsupervised}.
Consequently, SSR risks irreversibly discarding hard but correctly labeled samples near decision boundaries~\cite{arpit2017closer}, whereas NR-SEI's clustering struggles when class-irrelevant features dominate the signal distributions.
Second, by neglecting self-supervised pre-training, they fail to establish a label-independent feature space~\cite{he2020momentum, chen2020simple}, which is crucial for unbiased noise identification.
Finally, the absence of a recovery mechanism for inadvertently filtered samples leads to the suboptimal utilization of valuable training data.
These limitations highlight the necessity of shifting towards self-supervised learning and iterative sample recovery.

To intuitively illustrate the evolution of label-noise robust SEI, we summarize the existing paradigms in Fig. 1.
Traditional DL-based SEI methods (Fig. 1(a)) rigidly operate under a clean-label assumption, rendering them highly vulnerable to training data with label noise.
To mitigate this, recent approaches (Fig. 1(b)) introduce post-training sample selection.
However, because their feature extraction relies heavily on supervised signals, the learned representations are inevitably contaminated by the very noisy labels they aim to filter, creating a detrimental confirmation bias.
To circumvent this compounding confirmation bias, a new paradigm (Fig. 1(c)) is required—one that strictly decouples representation learning from noise detection to fundamentally bypass the feature contamination bottleneck that plagues existing methods~\cite{li2022selective}.
While traditional unsupervised reconstructive methods (e.g., autoencoders) often waste representational capacity on fitting dynamic channel noise, contrastive learning explicitly forces the model to ignore these channel-induced variations~\cite{11052286, chen2020simple}.
By maximizing the agreement between RF-augmented views, it naturally extracts highly discriminative, hardware-intrinsic representations without requiring any label priors.
Motivated by this insight, we propose SEI-SHIELD, a novel and robust SEI framework specifically designed to actualize this decoupled paradigm and thwart label noise attacks in RF environments.
First, to eliminate the reliance on untrustworthy supervised signals, SEI-SHIELD employs Momentum Contrast (MoCo)~\cite{he2020momentum} enhanced with augmentations.
This enables the model to learn intrinsically robust, label-independent representations directly from complex-valued I/Q signals.
Building upon this pristine feature space, we introduce a K-Nearest Neighbors (KNN)-based noise identification module~\cite{bahri2020deep}, which isolates corrupted samples by evaluating neighborhood label consistency.
Furthermore, recognizing that simplistic filtering inevitably discards valuable hard samples situated near decision boundaries, we design an iterative sample rescue mechanism.
By meticulously evaluating prediction confidence alongside the feature-space similarity to class centroids (prototypes built from reliably clean signals)~\cite{snell2017prototypical}, this mechanism progressively recovers inadvertently discarded genuine samples, thereby maximizing the utilization of training data without compromising system security.
The contributions of our work are as follows.
\begin{enumerate}
\item[1)] 
We propose SEI-SHIELD, introducing a decoupled training paradigm that effectively circumvents the compounding confirmation bias inherent in existing label-noise learning methods. 
By leveraging self-supervised contrastive learning coupled with RF-tailored augmentations, our framework extracts highly discriminative, hardware-intrinsic representations from complex-valued I/Q signals, operating entirely independent of noisy label priors.
\item[2)] 
Building upon the label-independent latent space, we design a highly reliable label correction mechanism. 
Beyond standard KNN-based noise isolation, we introduce an innovative iterative rescue strategy that synergistically evaluates prediction confidence and feature-space similarity to class prototypes.
This mechanism safely recovers inadvertently discarded hard samples, maximizing training data utility even in severely poisoned environments.
\item[3)] 
We conduct extensive evaluations on two widely adopted real-world RF datasets (POWDER and ORACLE).
The results demonstrate that SEI-SHIELD establishes a new robust baseline, achieving state-of-the-art (SOTA) authentication accuracy under various noise rates and substantially outperforming existing regularization and sample-selection frameworks.
\end{enumerate}
The rest of this paper is organized as follows. Section II describes tbe related work. 
Section III introduces the preliminary theory.
Section IV details the proposed method.
Section V illustrates the experimental setup and analysis of the results.
Finally, Section VI presents the conclusion and future work.

\section{Related Work}
In this section, we review the existing works related to deep learning-based SEI and learning with noisy labels.

\subsection{Deep Learning for SEI}
Traditional SEI methods rely on hand-crafted features such as cyclostationary statistics, higher-order cumulants, and transient signal characteristics to distinguish individual emitters.
While effective under controlled conditions, these approaches require extensive domain expertise and are sensitive to varying channel environments.

The advent of deep learning has shifted the paradigm toward end-to-end feature extraction directly from raw in-phase and quadrature (I/Q) signal samples.
O'Shea et al.~\cite{o2016convolutional} pioneered the application of convolutional neural networks (CNNs) to RF signal classification, demonstrating that learned representations can surpass hand-crafted alternatives.
Merchant et al.~\cite{merchant2018deep} further showed that CNNs can learn discriminative RF fingerprints directly from raw I/Q data for device identification in cognitive radio networks.
Sankhe et al.~\cite{sankhe2019oracle} developed the ORACLE system, which achieves high-accuracy emitter identification over the air using deep CNNs.
Al-Shawabka et al.~\cite{al2020exposing} investigated the impact of wireless channel effects on RF fingerprinting reliability, revealing that channel variations can significantly degrade identification accuracy.
Shen et al.~\cite{shen2022towards} addressed scalability and channel robustness for LoRa device identification.
More recently, complex-valued neural networks (CVNNs)~\cite{wang2021efficient} have emerged as a natural fit for RF signal processing, as they preserve the inherent amplitude-phase coupling of I/Q signals that is discarded by real-valued alternatives.

Despite these advances, the aforementioned methods universally operate under the assumption that training labels are correct. 
In practical non-cooperative environments, however, this assumption often fails.
Delicate RFFs are easily distorted by dynamic wireless channels, which naturally introduces systemic ambiguities and human annotation errors.
Furthermore, the open nature of wireless networks exposes systems to adversarial data poisoning, where intelligent jammers deliberately inject misleading signals.
Consequently, this prevalent label noise poses a severe threat, fundamentally undermining the reliability of traditional SEI methods.

\subsection{Learning with Noisy Labels}
Learning with noisy labels has been extensively studied in the machine learning community, with existing methods broadly categorized into sample selection, robust loss design, and semi-supervised approaches.

Sample selection methods identify presumably clean samples and train exclusively on the selected subset.
Co-teaching\cite{han2018co} trains two networks simultaneously, where each network selects small-loss samples to teach the other, exploiting the disagreement between the two networks to filter noise.
Decoupling\cite{malach2017decoupling} further separates the decision of when to update from how to update, training only on samples where two networks disagree.
MentorNet\cite{jiang2018mentornet} learns a data-driven curriculum that down-weights noisy samples during student network training.
These methods share a common reliance on the small-loss criterion, which assumes clean samples incur smaller losses---an assumption that breaks down when noise rates are high or when hard samples naturally exhibit large losses.

Robust loss methods modify the training objective to reduce the impact of label noise.
Ghosh et al.\cite{ghosh2017robust} proved that the mean absolute error is theoretically noise-tolerant but suffers from underfitting in practice.
Generalized cross entropy (GCE)\cite{zhang2018generalized} interpolates between cross-entropy and MAE to balance robustness and learnability.
Label smoothing regularization\cite{szegedy2016rethinking} and Mixup\cite{zhang2018mixup} provide implicit robustness through regularization.
While these approaches improve tolerance to moderate noise, they do not explicitly identify corrupted samples, limiting their effectiveness under severe corruption.

Semi-supervised and hybrid approaches attempt to exploit noisy samples rather than simply discarding them.
DivideMix\cite{li2020dividemix} uses a Gaussian Mixture Model to divide training data into clean and noisy partitions and treats noisy samples as unlabeled data within a MixMatch framework.
Confident learning\cite{northcutt2021confident} estimates the joint distribution of noisy and true labels to systematically identify label errors.
More recently, contrastive learning has been integrated into the LNL pipeline: Sel-CL\cite{li2022selective} performs selective supervised contrastive learning by filtering noisy pairs, and UNICON\cite{karim2022unicon} combines uniform sample selection with contrastive regularization.
Although these hybrid methods represent the current SOTA on image benchmarks, they operate exclusively on real-valued data and cannot be directly transferred to complex-valued RF signals, where the amplitude-phase coupling carries discriminative information essential for emitter identification.

In the SEI domain, only a few works have directly addressed label noise.
SSR\cite{SSR-SEI} employs a two-stage adaptive sample selection strategy driven by confidence learning, combined with label smoothing and entropy minimization for regularization.
NR-SEI\cite{NR-SEI} utilizes Jensen-Shannon divergence as a global cleanliness measure and applies clustering-based partitioning for semi-supervised training.
However, as analyzed in Section~I, both methods rely on supervised signals that are already corrupted by noise.
This inevitably leads to confirmation bias. 
Furthermore, they permanently lose correctly labeled hard samples that are mistakenly discarded during the filtering process.

To overcome these limitations, feature extraction must be decoupled from noisy labels, and self-supervised contrastive learning offers a natural solution.
By learning representations entirely from unlabeled data, it prevents the model from memorizing noisy patterns, effectively breaking the confirmation bias cycle. 
Building on this, our framework leverages these label-independent representations specifically for robust noise identification and iterative sample rescue.

\section{Preliminary Theory}
In this section, we formulate the mathematical model of RFFs and define the label noise problem in practical non-cooperative SEI scenarios.

\subsection{Radio Frequency Fingerprints Modeling}
SEI serves as a physical-layer authentication technique that distinguishes wireless devices based on their inherent hardware imperfections.
These unique imperfections, collectively known as RFFs, are introduced unintentionally during the manufacturing process of the transmitter chain components, such as the Digital-to-Analog Converter (DAC), mixers, local oscillators, and Power Amplifiers (PA).

Following the signal transmission process, the received complex-valued RF signal can be mathematically modeled as:
\begin{equation}
r(t) = h(t) \cdot f_{\text{hw}}(x(t)) + n(t),
\end{equation}
where $x(t)$ is the ideal baseband modulation signal carrying the transmitted information, and $f_{\text{hw}}(\cdot)$ represents the composite nonlinear distortion introduced by the transmitter hardware chain (including DAC quantization errors, I/Q imbalance, oscillator phase noise, and PA nonlinearity).
Furthermore, $h(t)$ denotes the complex-valued wireless channel coefficient under the narrowband flat-fading assumption, and $n(t)$ represents the additive white Gaussian noise (AWGN).

The device-specific RFF is inherently embedded within $f_{\text{hw}}(\cdot)$, which varies across individual transmitters due to subtle manufacturing tolerances.
The fundamental goal of SEI is to extract and identify these hardware-intrinsic signatures from the received signal $r(t)$, despite the severe interference from the dynamic channel $h(t)$ and background noise $n(t)$.

In our digital processing pipeline, the continuous received signal $r(t)$ is sampled and digitized into a finite sequence $\mathbf{r} \in \mathbb{R}^{2 \times L}$, where $L$ is the sample length.
The two dimensions correspond to the In-phase (I) and Quadrature (Q) components, which jointly preserve the critical amplitude-phase coupling essential for identifying specific emitters.

\subsection{Label Noise Model in Non-Cooperative Environments}
In an ideal supervised learning paradigm, an SEI model is trained on a perfectly curated dataset $\mathcal{D}^* = \{(\mathbf{r}_i, y_i^*)\}_{i=1}^N$, where $\mathbf{r}_i \in \mathbb{R}^{2 \times L}$ is the $i$-th sampled signal, $y_i^* \in \mathcal{Y} = \{1, 2, \dots, C\}$ is the ground-truth emitter label, and $C$ is the total number of emitter classes.
The objective is to learn a mapping function that correctly predicts the true label.

However, in practical non-cooperative spectrum monitoring or electronic warfare scenarios, obtaining perfectly annotated data is exceedingly difficult.
Signal labels are often corrupted due to severe channel fading-induced ambiguities, alignment errors in the automated signal interception pipeline, or adversarial label poisoning attacks by intelligent jammers seeking to evade detection.
Consequently, the actual available training dataset is a noisy version, denoted as $\mathcal{D} = \{(\mathbf{r}_i, \tilde{y}_i)\}_{i=1}^N$, where $\tilde{y}_i$ is the observed noisy label that may differ from $y_i^*$.

To quantify the degree and structure of this label corruption, we introduce the noise transition matrix $T \in [0, 1]^{C \times C}$.
Each element $T_{jk}$ denotes the probability that a true label $j$ is corrupted and observed as label $k$:
\begin{equation}
T_{jk} = P(\tilde{y} = k \mid y^* = j).
\end{equation}

In this work, we primarily focus on the symmetric noise model, which is a rigorous standard for evaluating algorithmic robustness.
It simulates a worst-case scenario where an attacker randomly perturbs the labels or human annotators make unbiased random mistakes.
Given an overall label noise rate $\eta \in \left[ 0, 1 \right)$, the transition probabilities for the symmetric noise model are defined as:
\begin{equation}
T_{jk} = 
\begin{cases} 
1 - \eta, & \text{if } j = k, \\
\frac{\eta}{C - 1}, & \text{if } j \neq k.
\end{cases}
\end{equation}
This indicates that a true label $j$ has a probability of $1 - \eta$ to remain correct, and a probability of $\eta$ to be uniformly flipped to any of the other $C - 1$ classes.
The overarching challenge for our proposed SEI-SHIELD framework is to extract reliable, hardware-intrinsic representations and optimize the SEI classifier using only this label-corrupted dataset $\mathcal{D}$, without relying on any prior knowledge of the noise transition matrix $T$ or access to a clean validation set.

\begin{figure*}[ht]
	\centering
	\includegraphics[width=\textwidth]{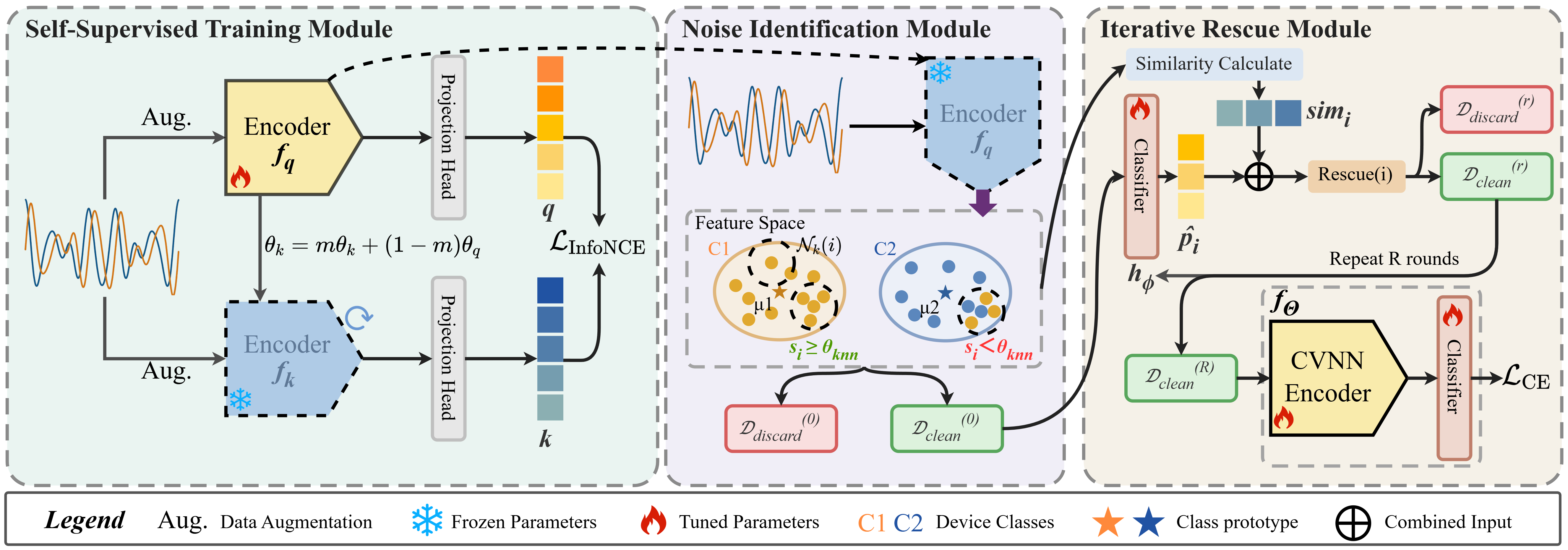}
	\caption{
    Framework of the proposed SEI-SHIELD.
    A self-supervised training module first learns label-independent signal representations from augmented raw I/Q samples by momentum contrastive learning.
    A KNN-based noise identification module then uses the frozen encoder to partition the training set into an initial clean subset $\mathcal{D}_{\text{clean}}^{(0)}$ and a discarded subset $\mathcal{D}_{\text{discard}}^{(0)}$ according to neighborhood label consistency in the learned feature space. 
    An iterative rescue module finally re-evaluates discarded samples with classifier confidence and prototype similarity, progressively recovering hard but correctly labeled samples over $R$ rounds to obtain the refined clean set used for final supervised SEI training.
    }
	\label{fig:framework}
\end{figure*}

\section{Methodology}
\subsection{Overview}
We propose SEI-SHIELD, a robust SEI framework consisting of four key modules, a self-supervised contrastive training module, a KNN-based noise identification module, an iterative hard sample rescue module, and an end-to-end CVNN classifier, as illustrated in Fig. 2.
The self-supervised training module employs Momentum Contrast (MoCo) enhanced with RF-tailored augmentations to learn intrinsically robust, label-independent representations directly from raw complex-valued I/Q signals, preventing the encoder from memorizing noisy label patterns and thereby fundamentally eliminating the confirmation bias inherent in existing supervised approaches.
Besides, the noise identification module leverages these uncontaminated representations to partition the training set into a provisionally clean subset $\mathcal{D}_{\text{clean}}^{(0)}$ and a discarded subset $\mathcal{D}_{\text{discard}}^{(0)}$ by evaluating the neighborhood label consistency of each sample via K-nearest neighbors analysis.
Moreover, recognizing that one-pass filtering inevitably misclassifies hard but correctly labeled samples near decision boundaries as noise, the iterative rescue module progressively recovers these valuable samples by jointly leveraging prediction confidence and prototype-based cosine similarity.
Finally, the refined clean dataset $\mathcal{D}_{\text{clean}}^{(R)}$ is used to train a CVNN in an end-to-end manner, preserving the inherent amplitude-phase coupling of I/Q signals for SEI.

\subsection{Self-Supervised Contrastive Training}
As discussed in Section I, existing noise-robust SEI methods rely on supervised signals that are already corrupted by label noise, leading to confirmation bias in the learned feature space.
To fundamentally avoid this issue, SEI-SHIELD decouples the representation learning stage from noisy label supervision entirely by employing self-supervised contrastive training.
Specifically, we adopt Momentum Contrast (MoCo)~\cite{he2020momentum} as the contrastive learning framework and design a set of RF-tailored augmentations for generating positive sample pairs from raw I/Q signals.

\subsubsection{RF-Tailored Data Augmentation}
The data augmentation module is designed to generate augmented views that preserve hardware-intrinsic RFFs while introducing realistic variations in channel-dependent characteristics.
Unlike image-based contrastive learning that commonly employs random cropping and color jittering~\cite{chen2020simple}, RF signals require domain-specific augmentations that simulate the physical phenomena encountered in wireless propagation environments.
To this end, we design seven augmentation operations specifically tailored for I/Q signals, categorized into three groups based on the type of variation they simulate.

The first group targets amplitude-domain variations to model the fading and gain fluctuations encountered in practical wireless channels.
Amplitude scaling multiplies the signal by a random factor $\epsilon \sim \mathcal{N}(1, \sigma_s^2)$, simulating flat fading.
Magnitude warping applies a smooth, time-varying amplitude envelope generated via cubic spline interpolation over $K_m + 2$ control points sampled from $\mathcal{N}(1, \sigma_m^2)$, emulating frequency-selective fading effects.
Random sign flip with channel permutation simulates the phase ambiguity and I/Q swap that may arise from receiver imperfections.

The second group addresses temporal-domain distortions to simulate timing offsets and multipath-induced effects.
Temporal permutation divides the time axis into $S$ equal segments, where $S \sim \text{Uniform}(1, S_{\max})$, and randomly shuffles their order, disrupting the temporal structure while preserving spectral content.
Time warping constructs a smooth nonlinear time mapping via cubic spline interpolation with $K_t + 2$ control points and resamples the signal accordingly.

The third group comprises window-based augmentations that simulate partial signal observation and local propagation anomalies.
Window slicing extracts a contiguous sub-window of length $\lceil \rho_s L \rceil$ and resamples it to the original length $L$ via linear interpolation.
Window warping selects a local region of length $\lceil \rho_w L \rceil$, stretches or compresses it by a randomly selected scale factor, and resamples the full signal back to $L$.

For each training sample $\mathbf{r}_i \in \mathbb{R}^{2 \times L}$, two augmented views $\mathbf{r}_i^{(1)}$ and $\mathbf{r}_i^{(2)}$ are generated independently.
Each view is produced by sequentially applying $n_{\text{aug}}$ randomly selected operations without replacement from the pool of seven methods, where $n_{\text{aug}} \sim \text{Uniform}(2, 4)$.
The independent selection ensures that the two views undergo distinct transformation pipelines, forcing the encoder to learn representations invariant to all channel-induced variations while retaining the discriminative hardware fingerprints.

\subsubsection{Complex-Valued Backbone Encoder}
The backbone encoder $f_\theta$ is built upon a CVNN architecture that natively preserves the amplitude-phase coupling of I/Q signals.
The input signal $\mathbf{r}_i \in \mathbb{R}^{2 \times L}$ is treated as a single complex-valued channel, where the two dimensions correspond to the real (in-phase) and imaginary (quadrature) components.

Each complex convolution layer implements the algebraic rule for complex-valued multiplication.
Given real-valued convolution kernels $W_{\text{re}}$ and $W_{\text{im}}$, the complex convolution is defined as:
\begin{equation}
\begin{aligned}
\text{Re}(\mathbf{y}) &= W_{\text{re}} * \text{Re}(\mathbf{x}) - W_{\text{im}}
* \text{Im}(\mathbf{x}), \\
\text{Im}(\mathbf{y}) &= W_{\text{re}} * \text{Im}(\mathbf{x}) + W_{\text{im}}
* \text{Re}(\mathbf{x}),
\end{aligned}
\end{equation}
where $*$ denotes the standard 1-D convolution operator.
The real and imaginary outputs are concatenated along the channel dimension, yielding a feature map with $2 \times C_{\text{out}}$ real-valued channels for $C_{\text{out}}$ complex-valued filters.

The encoder comprises $N_b$ cascaded complex convolutional blocks, each consisting of a complex convolution with $C_{\text{out}}$ filters, followed by ReLU activation, batch normalization over the $2C_{\text{out}}$ concatenated channels, and max pooling.
A flatten operation and a fully connected layer then produce a $d_e$-dimensional embedding, activated by ReLU:
\begin{equation}
\mathbf{z}_i = f_\theta(\mathbf{r}_i) \in \mathbb{R}^{d_e}.
\end{equation}

Compared with real-valued alternatives that treat the I and Q components as independent channels, the complex convolution explicitly models their interdependence through the cross terms $W_{\text{im}} * \text{Im}(\mathbf{x})$ and $W_{\text{im}} * \text{Re}(\mathbf{x})$, thereby capturing the joint amplitude-phase distortions introduced by transmitter hardware imperfections.

\subsubsection{Momentum Contrast Framework}
Given the augmented view pairs $\{(\mathbf{r}_i^{(1)}, \mathbf{r}_i^{(2)})\}_{i=1}^N$, MoCo trains a query encoder $f_q$ and a momentum-updated key encoder $f_k$ to learn discriminative representations 
through contrastive learning.
Both encoders share the identical CVNN backbone described above, each followed by a projection head that maps the $d_e$-dimensional embedding into a lower-dimensional space through a three-layer MLP:
\begin{equation}
g(\mathbf{z}) = W_3 \, \sigma(W_2 \, \sigma(W_1 \mathbf{z})),
\end{equation}
where $W_1 \in \mathbb{R}^{d_p \times d_e}$, $W_2 \in \mathbb{R}^{(d/4) \times d_p}$, $W_3 \in \mathbb{R}^{d \times (d/4)}$, $d_p$ is the hidden dimension, $d$ is the projection dimension, and $\sigma(\cdot)$ denotes the composition of batch normalization and ReLU activation.

For a mini-batch of $N$ samples, the query encoder processes the first views to produce $\ell_2$-normalized query vectors $\mathbf{q}_i = g_q(f_q(\mathbf{r}_i^{(1)})) / \|g_q(f_q(\mathbf{r}_i^{(1)}))\|_2$.
The key encoder processes the second views to produce key vectors $\mathbf{k}_i$ in the same manner, but without gradient computation.
The key encoder parameters $\theta_k$ are updated via exponential moving average (EMA) of the query encoder parameters $\theta_q$:
\begin{equation}
\theta_k \leftarrow m \cdot \theta_k + (1 - m) \cdot \theta_q,
\end{equation}
where $m$ is the momentum coefficient.
This slow-moving update ensures a consistent and smooth evolution of the key representations, stabilizing the contrastive learning process.

The contrastive objective treats each query--key pair $(\mathbf{q}_i, \mathbf{k}_i)$ derived from the same sample as a positive pair.
The negative samples are drawn from a first-in-first-out queue of size $K$ that accumulates $\ell_2$-normalized key embeddings from preceding mini-batches, providing a large and consistent negative set without requiring a large batch size.
The model is then optimized using the InfoNCE loss~\cite{oord2018representation}:
\begin{equation}
\mathcal{L}_{\text{InfoNCE}} = -\frac{1}{N}\sum_{i=1}^{N} \log
\frac{\exp(\mathbf{q}_i \cdot \mathbf{k}_i / \tau)}{\exp(\mathbf{q}_i \cdot
\mathbf{k}_i / \tau) + \sum_{j=1}^{K} \exp(\mathbf{q}_i \cdot \mathbf{k}_j^{-}
/ \tau)},
\end{equation}
where $\mathbf{k}_j^{-}$ denotes the $j$-th negative key retrieved from the queue, and $\tau$ is the temperature parameter that controls the sharpness of the similarity distribution.

\subsubsection{Feature Extraction}
After pre-training, only the backbone of the query encoder $f_q$ is retained, and the projection head $g(\cdot)$ is discarded, as the projection space is optimized for the contrastive objective rather than downstream discrimination~\cite{chen2020simple}.
All encoder parameters are frozen, and $d_e$-dimensional embeddings are extracted for the entire training set:
\begin{equation}
\mathbf{z}_i = f_q(\mathbf{r}_i), \quad \hat{\mathbf{z}}_i =
\frac{\mathbf{z}_i}{\|\mathbf{z}_i\|_2 + \epsilon}, \quad i = 1, \ldots, N,
\end{equation}
where $\epsilon$ is a small constant for numerical stability.
The resulting $\ell_2$-normalized embeddings $\{\hat{\mathbf{z}}_i\}_{i=1}^N$ reside on the unit hypersphere $\mathcal{S}^{d_e - 1}$, where the cosine similarity between any two samples directly corresponds to their inner product.
Since the entire pre-training stage operates without any label information, the learned feature space is entirely free from the contamination of noisy labels, establishing an unbiased foundation for the subsequent noise identification and sample rescue stages.

\subsection{KNN-Based Noise Identification}
Building upon the label-independent feature space established by the self-supervised training stage, we now address the problem of identifying which training samples carry corrupted labels.
The key insight is that, in an uncontaminated feature space, samples sharing the same true emitter identity tend to cluster together regardless of their observed labels.
Consequently, a clean sample whose observed label matches its true identity will find that a high proportion of its nearest neighbors share the same label, whereas a mislabeled sample will exhibit inconsistency with its neighborhood.

To quantify this property, we define the neighborhood label consistency score for each sample $i$.
Given the $\ell_2$-normalized embeddings $\{\hat{\mathbf{z}}_i\}_{i=1}^N$ on the unit hypersphere, we retrieve the set of $k$ nearest neighbors $\mathcal{N}_k(i)$ based on cosine similarity, and compute:
\begin{equation}
s_i = \frac{1}{k}\sum_{j \in \mathcal{N}_k(i)} \mathds{1}[\tilde{y}_j =
\tilde{y}_i],
\end{equation}
where $\tilde{y}_i$ is the observed noisy label and $\mathds{1}[\cdot]$ is the indicator function.
Intuitively, $s_i$ approaches 1 when the local neighborhood unanimously agrees with the observed label of sample $i$, and decreases toward 0 when the neighborhood is dominated by samples from other classes, signaling a likely label corruption.

The training set is then partitioned into a provisionally clean subset and a discarded subset based on a threshold $\theta_{knn}$:
\begin{equation}
\mathcal{D}_{\text{clean}}^{(0)} = \{(\mathbf{r}_i, \tilde{y}_i) \mid s_i \geq
\theta_{knn}\},
\end{equation}
\begin{equation}
\quad \mathcal{D}_{\text{discard}}^{(0)} = \{(\mathbf{r}_i,
\tilde{y}_i) \mid s_i < \theta_{knn}\},
\end{equation}

In practice, aggressive thresholding may leave certain classes severely underrepresented in $\mathcal{D}_{\text{clean}}^{(0)}$, which would impair the training of the classifier.
To prevent this, we introduce a class-level floor guarantee: for each class $c$, if the number of retained clean samples falls below a minimum count $n_{\min}$, the discarded samples belonging to class $c$ are sorted by their consistency scores $s_i$ in descending order, and the top-ranked ones are restored to $\mathcal{D}_{\text{clean}}^{(0)}$ until the count reaches $n_{\min}$.
This ensures that every class maintains sufficient representation for reliable classifier training while preserving the overall filtering quality.

It is worth emphasizing that the effectiveness of this module stems directly from the decoupled paradigm of SEI-SHIELD.
Because the KNN operates on self-supervised representations that have never been exposed to label information, the neighborhood structure faithfully reflects the signal similarity rather than the potentially erroneous label assignments, rendering the noise identification inherently immune to confirmation bias.

\subsection{Iterative Hard Sample Rescue}
Although the KNN-based noise identification effectively isolates the majority of corrupted samples, a fundamental trade-off persists: conservative thresholding retains noise, while aggressive thresholding inevitably discards correctly labeled hard samples that reside near decision boundaries.
These hard samples, despite carrying legitimate labels, exhibit low neighborhood consistency scores because their feature representations overlap with adjacent classes in the embedding space.
Permanently discarding them leads to an irreversible loss of valuable training data, particularly detrimental for underrepresented classes or closely spaced emitter signatures.

To recover these inadvertently filtered samples, we design an iterative rescue mechanism that progressively expands the clean subset $\mathcal{D}_{\text{clean}}$ by jointly evaluating prediction confidence and feature-space proximity to class prototypes.
The key insight is that a correctly labeled hard sample, although ambiguous in feature space, will nonetheless receive consistent support from both a classifier trained on verified clean data and the geometric structure of its true class cluster.
By iterating this evaluation over multiple rounds, the growing clean set yields increasingly representative prototypes and a more accurate classifier, enabling the rescue of progressively more challenging samples.

\subsubsection{Lightweight Classifier Training}
At the beginning of each rescue round $r \in \{1, 2, \ldots, R\}$, a lightweight linear classifier $h_\phi$ is trained on the current clean subset $\mathcal{D}_{\text{clean}}^{(r-1)}$ using the frozen self-supervised features as input.
Formally, for each sample $i$ in $\mathcal{D}_{\text{clean}}^{(r-1)}$, the classifier maps the $\ell_2$-normalized embedding $\hat{\mathbf{z}}_i$ to a probability distribution over $C$ classes:
\begin{equation}
\mathbf{p}_i = \text{softmax}(h_\phi(\hat{\mathbf{z}}_i)) \in \mathbb{R}^C.
\end{equation}
Training exclusively on verified clean samples ensures that the classifier's predictions are not biased by noisy labels, thereby providing a reliable confidence signal for the subsequent rescue decision.

\subsubsection{Class Prototype Construction}
To capture the geometric structure of each emitter class, we construct class prototypes from the current clean subset.
For each class $c \in \{1, 2, \ldots, C\}$, the prototype $\boldsymbol{\mu}_c^{(r)}$ is computed as the $\ell_2$-normalized centroid of all clean features belonging to class $c$:
\begin{equation}
\boldsymbol{\mu}_c^{(r)} = \frac{\bar{\mathbf{z}}_c}{\|\bar{\mathbf{z}}_c\|_2}, \quad \bar{\mathbf{z}}_c =
\frac{1}{|\mathcal{D}_c^{(r-1)}|} \sum_{i \in \mathcal{D}_c^{(r-1)}} \hat{\mathbf{z}}_i,
\end{equation}
where $\mathcal{D}_c^{(r-1)} = \{i \mid (i \in \mathcal{D}_{\text{clean}}^{(r-1)}) \wedge (\tilde{y}_i = c)\}$ denotes the set of clean samples labeled as class $c$.

\subsubsection{Sample Rescue via Joint Evaluation}
For each discarded sample $i \in \mathcal{D}_{\text{discard}}^{(r-1)}$, we compute two complementary rescue signals using the classifier $h_\phi$ trained in the current round and the class prototypes $\{\boldsymbol{\mu}_c^{(r)}\}_{c=1}^C$.

The first signal is the prediction confidence.
The classifier produces a softmax probability vector $\mathbf{p}_i$ for each discarded sample, from which we extract the maximum class probability and the predicted label:
\begin{equation}
\hat{p}_i = \max_{c} \, p_{i,c}, \quad \hat{y}_i = \arg\max_{c} \, p_{i,c}.
\end{equation}

The second signal is the prototype cosine similarity.
We measure the alignment between the sample's $\ell_2$-normalized embedding $\hat{\mathbf{z}}_i$ and the prototype of its predicted class:
\begin{equation}
\text{sim}_i = \hat{\mathbf{z}}_i \cdot \boldsymbol{\mu}_{\hat{y}_i}^{(r)}.
\end{equation}
Since both vectors reside on the unit hypersphere, this inner product directly yields the cosine similarity, quantifying how geometrically close the sample is to the centroid of its predicted class cluster.

A discarded sample is rescued and returned to $\mathcal{D}_{\text{clean}}^{(r)}$ if and only if two conditions are jointly satisfied: (i) the predicted label agrees with the observed noisy label, i.e., $\hat{y}_i = \tilde{y}_i$, ensuring label consistency; and (ii) the sample passes at least one of the following criteria:
\begin{equation}
\text{Rescue}(i) =
\begin{cases}
\text{True}, & \text{if } \hat{y}_i = \tilde{y}_i \text{ and } \hat{p}_i \geq \theta_{\text{high}}, \\
\text{True}, & \text{if } \hat{y}_i = \tilde{y}_i \text{ and } \hat{p}_i \geq \theta_{\text{low}} \text{ and } \text{sim}_i \geq \theta_{\text{sim}}, \\
\text{False}, & \text{otherwise},
\end{cases}
\end{equation}
where $\theta_{\text{high}}$, $\theta_{\text{low}}$, and $\theta_{\text{sim}}$ are the high-confidence threshold, low-confidence threshold, and similarity threshold, respectively.

The first criterion captures samples for which the classifier is highly confident, indicating strong discriminative evidence despite their initial filtering.
The second criterion recovers samples with moderate confidence but strong geometric affinity to their predicted class prototype, capturing hard samples that lie near decision boundaries yet clearly belong to the correct class cluster.
The mandatory label-match condition $\hat{y}_i = \tilde{y}_i$ serves as a safety gate, preventing the rescue of samples whose observed labels contradict the classifier's prediction, thereby avoiding the reintroduction of genuinely corrupted labels.

After each rescue round, the clean subset is updated as:
\begin{equation}
\mathcal{D}_{\text{clean}}^{(r)} = \mathcal{D}_{\text{clean}}^{(r-1)} \cup \{(\mathbf{r}_i, \tilde{y}_i) \mid \text{Rescue}(i) = \text{True}\}.
\end{equation}
The discard set is correspondingly reduced: $\mathcal{D}_{\text{discard}}^{(r)} = \mathcal{D}_{\text{discard}}^{(r-1)} \setminus \mathcal{D}_{\text{clean}}^{(r)}$.
This process repeats for $R$ rounds, during which the progressively expanding clean set yields more representative prototypes and a more accurate classifier, enabling the rescue of increasingly challenging samples in subsequent iterations.

After $R$ rounds of iterative rescue, the refined clean dataset $\mathcal{D}_{\text{clean}}^{(R)}$ is used to train the final CVNN classifier in an end-to-end manner using the standard cross-entropy loss.
The detailed network architecture and training configurations are provided in Section~V.

\subsection{Overall Training Procedure}
The complete SEI-SHIELD training pipeline proceeds in four sequential stages, as summarized in Algorithm~1.
First, the MoCo-based self-supervised module learns label-independent representations from raw I/Q signals using RF-tailored augmentations (Section~IV-B).
Then, the KNN-based noise identification module partitions the training set into provisionally clean and discarded subsets based on neighborhood label consistency (Section~IV-C).
Next, the iterative rescue mechanism progressively recovers correctly labeled hard samples from the discarded set over $R$ rounds by jointly evaluating prediction confidence and prototype similarity (Section~IV-D).
Finally, the refined clean dataset $\mathcal{D}_{\text{clean}}^{(R)}$ is used to train a CVNN classifier end-to-end on the raw I/Q signals with the standard cross-entropy loss, whose detailed architecture and training configurations are provided in Section~V.

\begin{algorithm}[t]
\caption{SEI-SHIELD Training Pipeline}
\label{alg:sei_shield}
\begin{algorithmic}[1]
\renewcommand{\algorithmicrequire}{\textbf{Input:}}
\renewcommand{\algorithmicensure}{\textbf{Output:}}
\REQUIRE Noisy training set $\mathcal{D} = \{(\mathbf{r}_i, \tilde{y}_i)\}_{i=1}^N$; KNN threshold $\theta_{knn}$; rescue thresholds $\theta_{\text{high}}$, $\theta_{\text{low}}$, $\theta_{\text{sim}}$; rescue rounds $R$.
\ENSURE Trained CVNN classifier $f_{\Theta^*}$.
\STATE \textit{// Stage 1: Self-Supervised Contrastive Pre-training}
\STATE Train MoCo with RF-tailored augmentations on $\{\mathbf{r}_i\}_{i=1}^N$.
\STATE Extract $\ell_2$-normalized features $\{\hat{\mathbf{z}}_i\}_{i=1}^N$ via frozen encoder $f_q$.
\STATE \textit{// Stage 2: KNN-Based Noise Identification}
\FOR{each sample $i = 1, \ldots, N$}
    \STATE Compute neighborhood label consistency $s_i$ via Eq.~(10).
\ENDFOR
\STATE Partition: $\mathcal{D}_{\text{clean}}^{(0)} \gets \{i \mid s_i \geq \theta_{knn}\}$, $\mathcal{D}_{\text{discard}}^{(0)} \gets \{i \mid s_i < \theta_{knn}\}$.
\STATE Apply class-level floor guarantee to $\mathcal{D}_{\text{clean}}^{(0)}$.
\STATE \textit{// Stage 3: Iterative Hard Sample Rescue}
\FOR{$r = 1$ to $R$}
    \STATE Train lightweight classifier $h_\phi$ on $\mathcal{D}_{\text{clean}}^{(r-1)}$.
    \STATE Compute class prototypes $\{\boldsymbol{\mu}_c^{(r)}\}_{c=1}^C$ via Eq.~(14).
    \FOR{each $i \in \mathcal{D}_{\text{discard}}^{(r-1)}$}
        \STATE Compute $\hat{p}_i$, $\hat{y}_i$, and $\text{sim}_i$ via Eqs.~(15)--(16).
        \IF{$\text{Rescue}(i) = \text{True}$ (Eq.~(17))}
            \STATE Move sample $i$ from $\mathcal{D}_{\text{discard}}^{(r-1)}$ to $\mathcal{D}_{\text{clean}}^{(r)}$.
        \ENDIF
    \ENDFOR
\ENDFOR
\STATE Train CVNN classifier on $\mathcal{D}_{\text{clean}}^{(R)}$ with standard cross-entropy loss.
\RETURN Trained CVNN classifier $f_{\Theta^*}$.
\end{algorithmic}
\end{algorithm}

\section{Experimental Evaluation}
\subsection{Experimental Settings}
\subsubsection{Implementation Details}
The proposed SEI-SHIELD framework was implemented using PyTorch 1.8.1 on a server running Ubuntu 20.04.3 LTS, equipped with Intel(R) Xeon(R) Gold 6330 processor and one NVIDIA GeForce RTX 3080 GPU.
The number of cascaded complex convolutional blocks $N_b$ for the final CVNN classifier is adapted to the input signal length of each dataset: $N_b = 8$ for ORACLE ($L = 2{,}048$) and $N_b = 6$ for POWDER ($L = 512$).
The MoCo projection head employs a three-layer MLP with dimensions $d_e \rightarrow 4d_e \rightarrow d/4 \rightarrow d$, where each hidden layer is followed by batch normalization and ReLU activation.
The final CVNN classification head comprises a projector (Linear $d_e \rightarrow 256$, BatchNorm, ReLU, Dropout) followed by a linear layer mapping to $C$ classes.
All remaining hyperparameters are summarized in Table~\ref{tab:hyperparams}, with equation references indicating the corresponding formulation in Section~IV.
\begin{table}[!t]
\caption{Hyperparameter Settings of SEI-SHIELD}
\label{tab:hyperparams}
\centering
\renewcommand{\arraystretch}{1.1}
\setlength{\tabcolsep}{10pt}
\footnotesize
\begin{tabular}{@{}ccc@{}}
\toprule
\textbf{Stage} & \textbf{Parameter} & \textbf{Value} \\
\midrule
\multirow{7}{*}{\makecell[c]{Data\\Augment.}}
 & Scaling $\sigma_s$ & 0.1 \\
 & Mag. warp $\sigma_m$, $K_m$ & 0.2, 4 \\
 & Time warp $\sigma_t$, $K_t$ & 0.2, 4 \\
 & Max segments $S_{\max}$ & 4 \\
 & Window slice $\rho_s$ & 0.9 \\
 & Window warp $\rho_w$ & 0.1 \\
 & Aug. count $n_{\text{aug}}$ & Uniform(2, 4) \\
\midrule
\multirow{8}{*}{MoCo}
 & Filters $C_{\text{out}}$ & 64 \\
 & Embed. dim $d_e$ & 1,024 \\
 & Proj. dim $d$ & 64 \\
 & Momentum $m$ \scriptsize{Eq.~(7)} & 0.99 \\
 & Temp. $\tau$ \scriptsize{Eq.~(8)} & 0.03 \\
 & Queue size $K$ & 512 \\
 & Epochs / LR / Batch & 300 / 5e-4 / 256 \\
\midrule
\multirow{3}{*}{KNN}
 & Neighbors $k$ \scriptsize{Eq.~(10)} & 20 \\
 & Threshold $\theta_{knn}$ \scriptsize{Eq.~(11)} & 0.4 \\
 & Class floor $n_{\min}$ & 35 \\
\midrule
\multirow{3}{*}{Rescue}
 & Rounds $R$ & 3 \\
 & $\theta_{\text{high}}$/$\theta_{\text{low}}$/$\theta_{\text{sim}}$ \scriptsize{Eq.~(17)} & 0.6/0.4/0.8 \\
 & Clf. epochs / LR & 100 / 1e-3 \\
\midrule
\multirow{3}{*}{\makecell[c]{Final\\CVNN}}
 & Blocks $N_b$ & 8 / 6\textsuperscript{$\dagger$} \\
 & Dropout $p$ & 0.5 \\
 & Epochs / LR / Batch & 100 / 1e-3 / 256 \\
\bottomrule
\multicolumn{3}{@{}l}{\textsuperscript{$\dagger$}\scriptsize{ORACLE ($L\!=\!2048$) / POWDER ($L\!=\!512$).}}
\end{tabular}
\end{table}

\subsubsection{Dataset Description}
We evaluated the proposed SEI-SHIELD framework on two publicly available real-world RF datasets with distinct scales and signal characteristics, as summarized in Table~\ref{tab:datasets}.

\textit{ORACLE Dataset}~\cite{sankhe2019oracle}: Recordings of raw I/Q signals were collected from over-the-air transmissions using 16 USRP X310 software-defined radios (SDRs) transmitting WiFi frames.
The receiver SDR processed incoming signals at a 5~MS/s sampling rate with a center frequency of 2.45~GHz.
Each signal sample contains $L = 2{,}048$ I/Q points.
We randomly selected 1{,}500 samples per emitter via stratified sampling, yielding a total of 24{,}000 samples across $C = 16$ classes.

\textit{POWDER Dataset}~\cite{9348261}: 
Recordings of raw I/Q signals were collected from over-the-air transmissions using $C = 4$ USRP X310 SDRs transmitting WiFi, 4G, and 5G frames.
The incoming signals were processed by the SDR receiver at a center frequency of 2.685 GHz, utilizing sampling rates of 5 MS/s for WiFi and 7.68 MS/s for 4G and 5G.
Each signal sample contains $L = 512$ I/Q points.
We randomly selected 1{,}000 samples per emitter via stratified sampling, yielding a total of 4{,}000 samples across $C = 4$ classes.

\begin{table}[!t]
	\caption{Summary of the ORACLE and POWDER Datasets}
	\label{tab:datasets}
	\centering
	\renewcommand{\arraystretch}{1.2}
	\setlength{\tabcolsep}{8pt}
	\footnotesize
	\begin{tabular}{ccc}
		\toprule
		\textbf{Property} & \textbf{ORACLE}~\cite{sankhe2019oracle} & \textbf{POWDER}~\cite{9348261} \\
		\midrule
		Signal Type       & WiFi                                    & WiFi/4G/5G                     \\
		Classes           & 16                                      & 4                              \\
		Samples per Class & 1{,}500                                 & 1{,}000                        \\
		Total Samples     & 24{,}000                                & 4{,}000                        \\
		Length ($L$)      & 2{,}048                                 & 512                            \\
		\bottomrule
	\end{tabular}
\end{table}

For both datasets, the selected samples were split into training, validation, and test sets with a ratio of 60\%/20\%/20\% using stratified sampling to preserve the class distribution.
Following the standard protocol in the noisy label literature~\cite{SSR-SEI, NR-SEI}, symmetric label noise was adopted to simulate annotation errors.
For a given noise ratio $\eta \in \{0.0, 0.1, 0.2, 0.3, 0.4, 0.5, 0.6\}$, a fraction $\eta$ of the training samples were randomly selected and their labels were uniformly replaced with one of the remaining $C - 1$ classes.
A binary indicator $z_i \in \{0, 1\}$ was maintained for each training sample to record whether its label had been corrupted, serving exclusively for evaluation purposes and never exposed to the training pipeline.
The test sets remained clean throughout all experiments.

\subsubsection{Baseline Methods}
The proposed SEI-SHIELD method is evaluated against the following established robust training techniques.
To ensure a fair comparison, all methods are applied to the same I/Q-formatted datasets, use the same CVNN backbone architecture, and are evaluated under identical data splits and noise configurations.
A summary of the baselines is provided below.

\begin{enumerate}
\item \textbf{CE}: The standard cross-entropy loss function used in classification tasks, serving as the lower-bound baseline without any noise-handling mechanism.
\item \textbf{Mixup}~\cite{zhang2018mixup}: A data augmentation method that creates synthetic training examples by interpolating between pairs of samples and their labels, thereby improving generalization under noisy conditions.
\item \textbf{LSR}~\cite{szegedy2016rethinking}: Label Smoothing Regularization, which reduces model overconfidence by distributing a fraction of the probability mass from the ground-truth class to all other classes, mitigating overfitting to potentially incorrect hard labels.
\item \textbf{GCE}~\cite{zhang2018generalized}: Generalized Cross-Entropy, a robust loss function that blends the benefits of mean absolute error and cross-entropy to achieve noise tolerance through a tunable parameter $q$.
\item \textbf{DML}~\cite{9846906}: Deep Metric Learning, which utilizes center loss to minimize the distance between feature embeddings and their corresponding class centers, thereby learning compact and discriminative representations.
\item \textbf{SSR}~\cite{SSR-SEI}: A two-stage adaptive sample selection method specifically designed for SEI under label noise, which employs confidence learning for coarse-grained selection and regularization-inspired loss (LSR + entropy minimization) for fine-grained selection.
\end{enumerate}

\subsubsection{Evaluation Metrics}
Following prior SEI studies~\cite{SSR-SEI}, the overall classification accuracy on the clean test set is adopted as the primary evaluation metric.
All experiments are conducted with a fixed random seed to ensure reproducibility.

\subsection{Identification Performance}
To comprehensively evaluate the robustness of the proposed SEI-SHIELD framework, we compare its identification accuracy against all baseline methods across symmetric noise rates $\eta \in \{0.0, 0.1, 0.2, 0.3, 0.4, 0.5, 0.6\}$ on both the ORACLE and POWDER datasets.
It is important to note that, due to the substantial difference in signal length between the two datasets ($L = 2{,}048$ for ORACLE and $L = 512$ for POWDER), the final CVNN classifier employs $N_b = 8$ cascaded complex convolutional blocks for ORACLE and $N_b = 6$ for POWDER, as specified in Table~\ref{tab:hyperparams}.
All baseline methods adopt the same dataset-specific CVNN backbone to ensure a fair comparison.
The complete results are presented in Tables~\ref{tab:oracle_results} and~\ref{tab:powder_results}.

\begin{table*}[!t]
\caption{Identification Accuracy (\%) on the ORACLE Dataset ($N_b = 8$) Under Different Noise Rates}
\label{tab:oracle_results}
\centering
\renewcommand{\arraystretch}{1.15}
\setlength{\tabcolsep}{8pt}
\footnotesize
\begin{tabular}{cccccccc}
\toprule
\textbf{Method} & $\boldsymbol{\eta = 0.0}$ & $\boldsymbol{\eta = 0.1}$ & $\boldsymbol{\eta = 0.2}$ & $\boldsymbol{\eta = 0.3}$ & $\boldsymbol{\eta = 0.4}$ & $\boldsymbol{\eta = 0.5}$ & $\boldsymbol{\eta = 0.6}$ \\
\midrule
CE          & 99.69 & 78.31 & 77.65 & 77.15 & 60.48 & \underline{64.81} & 41.90 \\
Mixup       & 93.92 & 95.15 & 88.48 & \underline{77.56} & 71.17 & 54.27 & 39.21 \\
LSR         & \textbf{99.90} & \textbf{95.46} & 87.17 & 75.27 & \underline{72.27} & 49.25 & 45.15 \\
GCE         & 99.48 & \underline{93.77} & \underline{90.31} & 77.27 & 61.73 & 60.73 & \underline{51.02} \\
DML         & \underline{99.85} & 89.27 & 72.62 & 70.29 & 60.60 & 64.58 & 39.56 \\
SSR         & 73.06 & 68.10 & 64.58 & 46.95 & 40.58 & 36.14 & 30.33 \\
\textbf{SEI-SHIELD (ours)}   & 97.19 & \textbf{95.46} & \textbf{94.00} & \textbf{91.25} & \textbf{74.96} & \textbf{73.69} & \textbf{55.44} \\
\bottomrule
\end{tabular}
\end{table*}

\begin{table*}[!t]
\caption{Identification Accuracy (\%) on the POWDER Dataset ($N_b = 6$) Under Different Noise Rates}
\label{tab:powder_results}
\centering
\renewcommand{\arraystretch}{1.15}
\setlength{\tabcolsep}{8pt}
\footnotesize
\begin{tabular}{ccccccccc}
\toprule
\textbf{Method} & $\boldsymbol{\eta = 0.0}$ & $\boldsymbol{\eta = 0.1}$ & $\boldsymbol{\eta = 0.2}$ & $\boldsymbol{\eta = 0.3}$ & $\boldsymbol{\eta = 0.4}$ & $\boldsymbol{\eta = 0.5}$ & $\boldsymbol{\eta = 0.6}$ \\
\midrule
CE          & \textbf{98.25} & 86.88 & 67.75 & 77.62 & 62.62 & 24.88 & 40.25 \\
Mixup       & 96.62 & \underline{94.12} & 88.62 & 85.25 & 81.25 & 67.88 & 55.25 \\
LSR         & 94.38 & 89.50 & 88.12 & 84.38 & 74.88 & 71.00 & 25.00 \\
GCE         & 97.62 & 90.62 & 85.00 & 72.75 & 47.38 & 32.75 & 25.38 \\
DML         & \textbf{98.25} & 88.62 & 79.25 & 64.88 & 50.38 & 48.25 & 29.12 \\
SSR         & 93.75 & 93.87 & \underline{89.75} & \underline{88.00} & \underline{83.62} & \underline{74.62} & \underline{57.37} \\
\textbf{SEI-SHIELD (ours)}   & \underline{97.75} & \textbf{96.75} & \textbf{95.88} & \textbf{93.00} & \textbf{88.25} & \textbf{89.75} & \textbf{78.62} \\
\bottomrule
\end{tabular}
\end{table*}

As presented in Tables~\ref{tab:oracle_results} and~\ref{tab:powder_results}, we highlight the best result in \textbf{bold} and the second best with \underline{underline} for each noise rate.
Overall, SEI-SHIELD consistently achieves the highest identification accuracy under label noise ($\eta \geq 0.1$) across both datasets.
A particularly noteworthy trend is that the performance margin between SEI-SHIELD and the strongest baseline widens progressively as the noise rate increases, underscoring the fundamental robustness advantage conferred by the decoupled training paradigm.

On the ORACLE dataset (Table~\ref{tab:oracle_results}), which comprises $C = 16$ emitter classes with signal length $L = 2{,}048$, all baseline methods exhibit pronounced accuracy degradation as the noise rate escalates.
Under the clean-label condition ($\eta = 0.0$), SEI-SHIELD attains 97.19\%, which is slightly below LSR (99.90\%) and DML (99.85\%).
This marginal gap is expected, as the KNN-based filtering stage inevitably removes a small fraction of correctly labeled samples even when no label noise is present, resulting in a reduced training set.
However, this minor trade-off is quickly compensated once label noise is introduced.
At $\eta = 0.1$, SEI-SHIELD ties with LSR at the highest accuracy of 95.46\%, indicating that even minimal noise suffices to activate the advantage of label-independent representations.
As the noise rate increases to moderate levels ($\eta = 0.2$ and $0.3$), the advantage of SEI-SHIELD becomes increasingly decisive: it achieves 94.00\% and 91.25\%, surpassing the second-best methods GCE (90.31\%) and Mixup (77.56\%) by 3.69\% and 13.69\%, respectively.
This widening margin demonstrates that supervised baselines suffer from compounding confirmation bias as more labels are corrupted, while the label-independent feature space of SEI-SHIELD remains uncontaminated.
Under severe label corruption ($\eta = 0.4$ through $0.6$), SEI-SHIELD continues to maintain a clear lead, achieving 74.96\%, 73.69\%, and 55.44\%, respectively.
Notably, the strongest baseline shifts across noise rates---from LSR (72.27\%) at $\eta = 0.4$ to CE (64.81\%) at $\eta = 0.5$ and GCE (51.02\%) at $\eta = 0.6$---reflecting the instability of supervised methods whose optimal operating points vary unpredictably with different noise rates.
In contrast, SEI-SHIELD exhibits consistent superiority across all tested noise rates, confirming that the decoupled training paradigm provides a fundamentally more robust solution.

On the POWDER dataset (Table~\ref{tab:powder_results}), which features $C = 4$ emitter classes with signal length $L = 512$, SEI-SHIELD demonstrates even more pronounced superiority.
Under the clean-label condition ($\eta = 0.0$), SEI-SHIELD achieves 97.75\%, marginally below CE and DML (both 98.25\%), reflecting the same negligible filtering overhead observed on ORACLE.
For all noise-corrupted settings ($\eta = 0.1$ through $0.6$), SEI-SHIELD consistently attains the highest accuracy, and the advantage margin grows substantially with the noise rate: from 2.63\% at $\eta = 0.1$ (96.75\% vs.\ Mixup 94.12\%) to 6.13\% at $\eta = 0.2$ (95.88\% vs.\ SSR 89.75\%), and ultimately to 21.25\% at $\eta = 0.6$ (78.62\% vs.\ SSR 57.37\%).
Notably, several regularization-based baselines collapse catastrophically at high noise rates: GCE drops from 97.62\% ($\eta = 0.0$) to 25.38\% ($\eta = 0.6$), and LSR plummets from 94.38\% to 25.00\%, which is effectively random guessing for a 4-class problem.
In contrast, SEI-SHIELD degrades gracefully from 97.75\% to 78.62\%, retaining a substantial portion of its discriminative capability even under extreme corruption.

In summary, the experimental results on both datasets reveal a consistent pattern: the advantage of SEI-SHIELD over existing methods amplifies as the label noise intensifies.
This behavior can be attributed to the core design principle of SEI-SHIELD---decoupling representation learning from noisy label supervision via self-supervised contrastive training.
The resulting label-independent feature space provides an uncontaminated foundation for both KNN-based noise identification and iterative sample rescue, enabling robust noise filtering even when 60\% of the training labels are corrupted.
While this decoupled pipeline incurs a negligible accuracy reduction under clean-label conditions, this minor trade-off is vastly outweighed by the substantial robustness gains in noisy environments, which are precisely the challenging scenarios that SEI-SHIELD is designed to address.

\subsection{Visualization Analysis}
\begin{figure*}[!t]
	\centering
	\includegraphics[width=\textwidth]{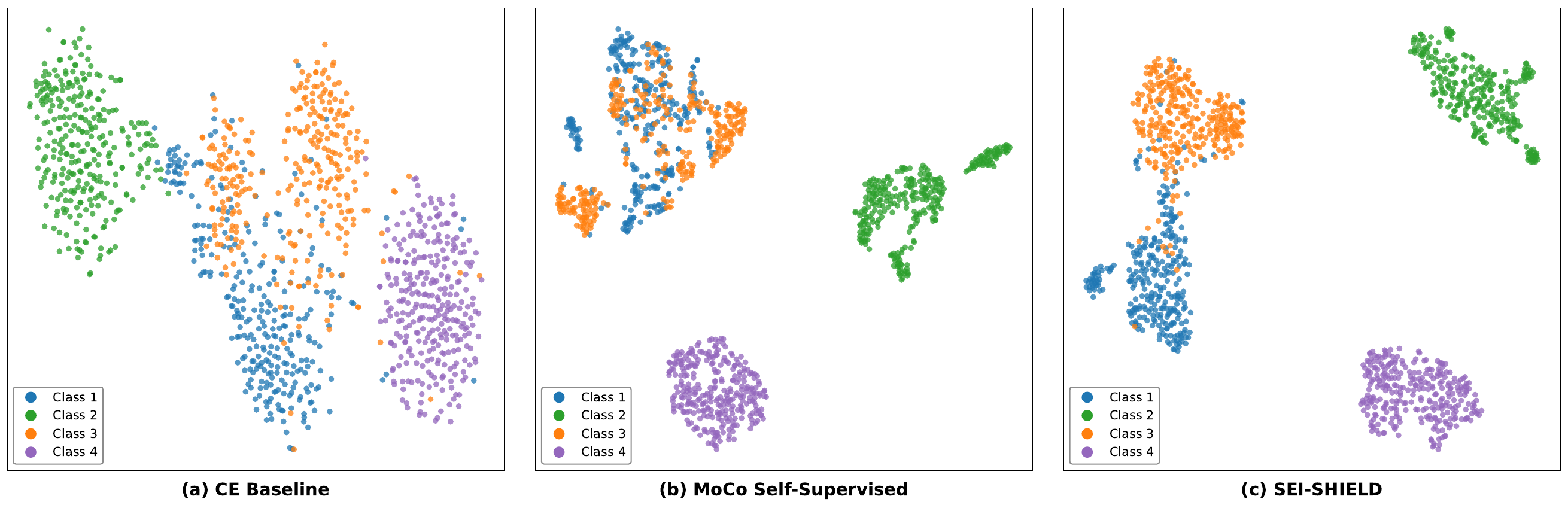}
	\caption{t-SNE visualization of learned feature representations on the POWDER dataset ($\eta = 0.3$). (a) CE Baseline trained with noisy labels exhibits severely entangled clusters due to feature contamination. (b) MoCo self-supervised training produces well-separated clusters without accessing any labels. (c) SEI-SHIELD achieves the most compact and discriminative clusters after noise filtering and iterative rescue.}
	\label{fig:tsne}
\end{figure*}

To provide intuitive insight into how the decoupled training paradigm establishes a label-independent feature space, we visualize the learned representations using t-SNE on the POWDER dataset under the noise rate $\eta = 0.3$.
Fig.~\ref{fig:tsne} compares the feature distributions of the supervised baseline against two key stages of our framework.

As shown in Fig.~\ref{fig:tsne}(a), the CE baseline trained with noisy labels achieves partial class separation; however, the inter-class distances remain small, and Class 1 (blue) and Class 3 (orange) exhibit severe overlap.
Additionally, Class 2 (green) slightly intersects with Class 1, and samples from other classes are scattered within the Class 4 (purple) region.
This limited separation arises from confirmation bias: the network memorizes erroneous label associations, learning features that conflate samples from distinct emitters.

Fig.~\ref{fig:tsne}(b) demonstrates that MoCo self-supervised training substantially improves the overall inter-class separation without accessing any labels.
Compared to the CE baseline, the spacing between clusters increases noticeably, and Class 2 and Class 4 become clearly distinguishable.
However, the overlap between Class 1 and Class 3 persists, indicating that certain emitter pairs share intrinsically similar RF characteristics that contrastive learning alone cannot fully disentangle.
Despite this limitation, the label-independent feature space provides a cleaner foundation for noise identification, as the learned representations are not biased by corrupted labels.

Finally, Fig.~\ref{fig:tsne}(c) shows that SEI-SHIELD produces the most compact and well-separated clusters among all three configurations.
After KNN-based noise identification and iterative rescue, the final classifier is trained on a refined dataset that excludes corrupted samples while preserving hard but correctly labeled ones.
This supervised training further sharpens the decision boundaries, eliminating most of the Class 1--3 overlap and yielding four reasonably distinct clusters.
The visualization corroborates the quantitative results in Table~\ref{tab:powder_results}, confirming that the proposed framework effectively addresses the feature contamination bottleneck inherent in purely supervised approaches.
\subsection{Ablation Study}
To evaluate the individual contribution of the KNN-based noise filtering module and the iterative rescue mechanism, we conduct an ablation study by systematically removing components from the SEI-SHIELD pipeline.
Three configurations are compared: (1) \textbf{CE}, the standard cross-entropy training without any noise-handling mechanism, serving as the unprotected baseline; (2) \textbf{w/o Rescue}, which retains the MoCo self-supervised pre-training and KNN noise filtering but omits the iterative rescue stage; and (3) \textbf{SEI-SHIELD}, the complete framework with all modules active.
The results on the ORACLE and POWDER datasets are presented in Tables~\ref{tab:ablation_oracle} and~\ref{tab:ablation_powder}, respectively.

\begin{table*}[!t]
	\caption{Ablation Study on the ORACLE Dataset ($C = 16$, $N_b = 8$): Identification Accuracy (\%) Under Different Noise Rates}
	\label{tab:ablation_oracle}
	\centering
	\renewcommand{\arraystretch}{1.15}
	\setlength{\tabcolsep}{8pt}
	\footnotesize
	\begin{tabular}{cccccccc}
		\toprule
		\textbf{Method}     & $\boldsymbol{\eta = 0.0}$ & $\boldsymbol{\eta = 0.1}$ & $\boldsymbol{\eta = 0.2}$ & $\boldsymbol{\eta = 0.3}$ & $\boldsymbol{\eta = 0.4}$ & $\boldsymbol{\eta = 0.5}$ & $\boldsymbol{\eta = 0.6}$ \\
		\midrule
		CE                  & \textbf{99.69}            & 78.31                     & 77.65                     & 77.15                     & 60.48                     & 64.81                     & 41.90                     \\
		w/o Rescue          & 64.85                     & 66.58                     & 64.10                     & 64.31                     & 54.77                     & 51.15                     & 48.38                     \\
		\textbf{SEI-SHIELD (ours)} & 97.19                     & \textbf{95.46}            & \textbf{94.00}            & \textbf{91.25}            & \textbf{74.96}            & \textbf{73.69}            & \textbf{55.44}            \\
		\bottomrule
	\end{tabular}
\end{table*}

\begin{table*}[!t]
	\caption{Ablation Study on the POWDER Dataset ($C = 4$, $N_b = 6$): Identification Accuracy (\%) Under Different Noise Rates}
	\label{tab:ablation_powder}
	\centering
	\renewcommand{\arraystretch}{1.15}
	\setlength{\tabcolsep}{8pt}
	\footnotesize
	\begin{tabular}{cccccccc}
		\toprule
		\textbf{Method}     & $\boldsymbol{\eta = 0.0}$ & $\boldsymbol{\eta = 0.1}$ & $\boldsymbol{\eta = 0.2}$ & $\boldsymbol{\eta = 0.3}$ & $\boldsymbol{\eta = 0.4}$ & $\boldsymbol{\eta = 0.5}$ & $\boldsymbol{\eta = 0.6}$ \\
		\midrule
		CE                  & \textbf{98.25}            & 86.88                     & 67.75                     & 77.62                     & 62.62                     & 24.88                     & 40.25                     \\
		w/o Rescue          & 92.12                     & 92.00                     & 90.75                     & 91.38                     & \textbf{90.62}                     & 87.50            & \textbf{83.50}            \\
		\textbf{SEI-SHIELD (ours)} & 97.75                     & \textbf{96.75}            & \textbf{95.88}            & \textbf{93.00}            & 88.25            & \textbf{89.75}                     & 78.62                     \\
		\bottomrule
	\end{tabular}
\end{table*}

On the ORACLE dataset (Table~\ref{tab:ablation_oracle}), removing the iterative rescue mechanism leads to a dramatic performance collapse.
The w/o Rescue configuration achieves only 64.85\% accuracy even under the clean-label condition ($\eta = 0.0$), which is substantially lower than both the CE baseline (99.69\%) and the full SEI-SHIELD framework (97.19\%).
This severe degradation can be attributed to the aggressive nature of KNN filtering in the 16-class setting: with a fixed threshold $\theta_{knn} = 0.4$, a considerable proportion of correctly labeled samples residing near inter-class decision boundaries exhibit low neighborhood consistency scores and are consequently discarded, leaving an insufficient training set for reliable classifier learning.
The iterative rescue mechanism addresses this critical limitation by progressively recovering these hard samples, yielding consistent improvements ranging from +7.06\% ($\eta = 0.6$) to +32.34\% ($\eta = 0.0$).

On the POWDER dataset (Table~\ref{tab:ablation_powder}), the interplay between the two modules reveals a more nuanced pattern that depends on the number of classes.
The w/o Rescue configuration already demonstrates strong noise robustness, maintaining 83.50\% accuracy even at $\eta = 0.6$---a level that surpasses all supervised baselines reported in Table~\ref{tab:powder_results}.
This indicates that, for a 4-class task with relatively well-separated emitter features, MoCo training combined with KNN filtering alone provides a sufficiently clean training subset.
The iterative rescue mechanism further improves accuracy at low-to-moderate noise rates ($\eta \leq 0.3$), with gains ranging from +1.62\% to +5.13\%.
However, a performance reversal is observed at $\eta = 0.4$ and $\eta = 0.6$: the w/o Rescue configuration outperforms SEI-SHIELD by 2.37\% at $\eta = 0.4$ and 4.88\% at $\eta = 0.6$.
This reversal can be attributed to the number of classes in the dataset.
With only $C = 4$ classes, a noisy sample has a probability of $1/(C-1) \approx 33.3\%$ of its predicted label coincidentally matching the erroneous observed label, thereby satisfying the condition in Eq.~(17).
In contrast, the same probability on ORACLE ($C = 16$) is only $1/15 \approx 6.7\%$, rendering the condition far more selective.
Consequently, at high noise rates on POWDER, the rescue mechanism inadvertently reintroduces a fraction of corrupted samples, degrading the final classifier.

These ablation results collectively reveal a characteristic of the iterative rescue mechanism that depends on the number of classes.
For multi-class scenarios such as ORACLE, where KNN filtering alone is overly aggressive and the label agreement probability for noisy samples is inherently low, the rescue mechanism serves as an indispensable recovery stage that yields improvements of up to 32\%.
Conversely, for datasets with few classes like POWDER under severe corruption, the false positive rate increases.
This demonstrates that the overall efficacy of the rescue mechanism is closely tied to the number of classes.

\subsection{Hyperparameter Sensitivity}
\begin{figure}[!t]
	\centering
	\includegraphics[width=\columnwidth]{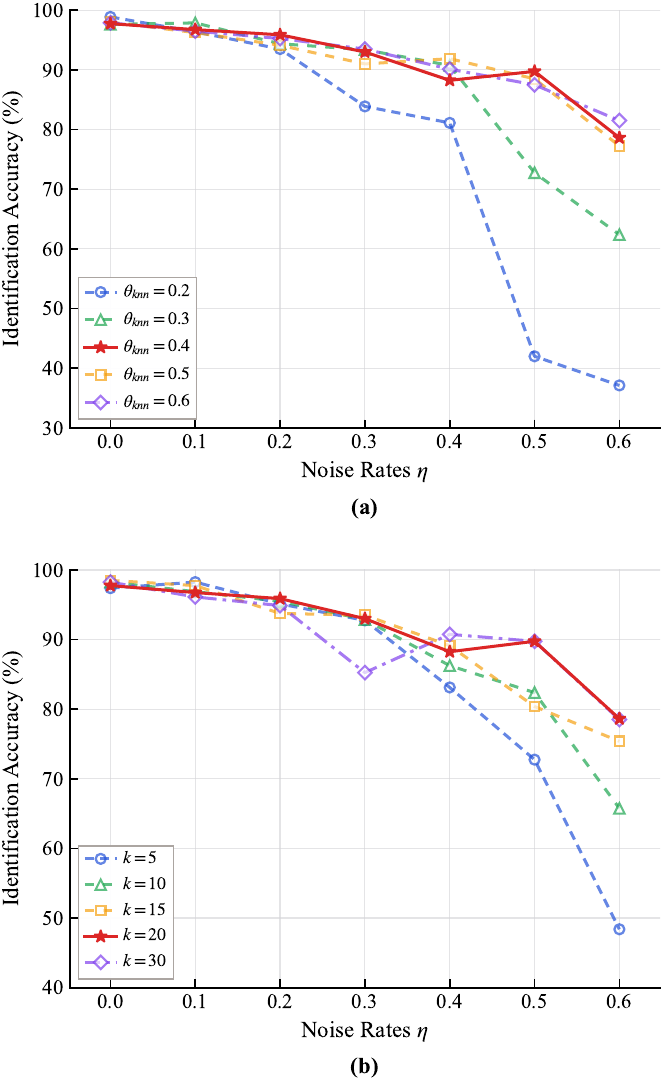}
	\caption{Hyperparameter sensitivity analysis on the POWDER dataset. (a) Effect of the KNN threshold $\theta_{knn}$ with $K = 20$ fixed. (b) Effect of the number of nearest neighbors $K$ with $\theta_{knn} = 0.4$ fixed. The red solid line with star markers indicates the default configuration ($\theta_{knn} = 0.4$, $K = 20$).}
	\label{fig:sensitivity}
\end{figure}

To assess the robustness of SEI-SHIELD to hyperparameter choices, we analyze the sensitivity of two critical parameters: the KNN threshold $\theta_{knn}$ and the number of nearest neighbors $K$.
Experiments are conducted on the POWDER dataset across the full range of noise rates $\eta \in \{0.0, 0.1, \ldots, 0.6\}$.
Fig.~\ref{fig:sensitivity} presents the identification accuracy as a function of $\eta$ for varying parameter values.

As shown in Fig.~\ref{fig:sensitivity}(a), the choice of $\theta_{knn}$ significantly impacts performance under high noise conditions.
When $\theta_{knn} = 0.2$, the framework achieves competitive accuracy at low noise rates ($\eta \leq 0.2$) but suffers a dramatic collapse at $\eta = 0.6$, dropping to approximately 37\%.
This degradation occurs because an overly permissive threshold admits a substantial proportion of corrupted samples into the clean set, undermining the subsequent classifier training.
Setting $\theta_{knn} = 0.3$ mitigates this issue but still exhibits noticeable degradation at $\eta \geq 0.5$.
In contrast, thresholds in the range $\theta_{knn} \in [0.4, 0.6]$ yield stable performance across all noise rates, maintaining accuracy above 77\% even at $\eta = 0.6$.
The marginal differences among these values indicate that the framework is robust to threshold selection within this range.

Fig.~\ref{fig:sensitivity}(b) reveals a similar pattern for the number of nearest neighbors $K$.
A small neighborhood size ($K = 5$) produces unstable noise identification: accuracy at $\eta = 0.6$ drops to approximately 48\%, as the consistency score becomes overly sensitive to local fluctuations in the feature space.
Increasing $K$ to 10 improves stability but still results in degraded performance at high noise rates.
For $K \geq 15$, the framework exhibits consistent behavior, with all configurations achieving approximately 76--79\% accuracy at $\eta = 0.6$.
Notably, $K = 20$ and $K = 30$ produce nearly identical results, suggesting that the consistency score stabilizes once a sufficient number of neighbors are considered.

These results demonstrate that SEI-SHIELD maintains robust performance across a reasonable range of hyperparameter values ($\theta_{knn} \in [0.4, 0.6]$, $K \in [15, 30]$).
The default configuration ($\theta_{knn} = 0.4$, $K = 20$) strikes a balance between noise filtering precision and sample retention, and practitioners can adjust these values within the identified stable ranges based on dataset characteristics without significant performance degradation.

\subsection{Computational Complexity Analysis}
To assess the feasibility of deploying SEI-SHIELD in resource-constrained edge IoT environments, we analyze its computational overhead in comparison with the CE baseline and the SSR method.
Table~\ref{tab:complexity} reports three complementary metrics: (1) offline total training time on a single NVIDIA RTX 3080 GPU, (2) single-sample inference FLOPs, and (3) average inference latency.
All measurements are conducted on the POWDER dataset.

\begin{table}[H]
	\caption{Computational Complexity Comparison on POWDER}
	\label{tab:complexity}
	\centering
	\renewcommand{\arraystretch}{1.2}
	\setlength{\tabcolsep}{4pt}
	\footnotesize
	\begin{tabular}{cccc}
		\toprule
		\textbf{Method} & \textbf{Train Time} & \textbf{Infer. FLOPs} & \textbf{Infer. Latency} \\
		\midrule
		CE              & 0.82 min            & 26.601 M              & 2.7400 ms                 \\
		SSR             & 0.93 min            & 26.601 M              & 2.3810 ms                 \\
		SEI-SHIELD      & 10.12 min           & 26.601 M              & 2.3831 ms                 \\
		\bottomrule
	\end{tabular}
\end{table}

As shown in Table~\ref{tab:complexity}, all three methods exhibit identical inference FLOPs (26.601 MFLOPs) and comparable inference latency (2--3 ms per sample).
This equivalence arises because the additional modules introduced by SEI-SHIELD, including the MoCo encoder, KNN noise filter, and iterative rescue mechanism, operate exclusively during offline training and are entirely stripped before deployment.
The actual inference pipeline deploys only the final CVNN backbone, which is architecturally identical across all methods.
Consequently, SEI-SHIELD imposes no additional computational burden at inference time, fully satisfying the real-time constraints of edge IoT devices.

The training time of SEI-SHIELD (10.12 min) is approximately $10\times$ longer than the CE baseline (0.82 min), primarily due to the computational overhead of self-supervised contrastive training and the iterative rescue procedure.
However, in practical non-cooperative SEI systems, such as base stations and spectrum monitoring centers, model training is typically performed offline on cloud servers or high-performance computing clusters, where training time is not a critical bottleneck.
The one-time offline investment of approximately 10 additional minutes yields accuracy improvements of up to 20\% under extreme 60\% label noise (Table~\ref{tab:powder_results}), representing a highly favorable trade-off for security-critical applications.

\section{Conclusion}
In this paper, we proposed SEI-SHIELD, a robust Specific Emitter Identification framework that addresses the critical challenge of label noise in physical-layer authentication systems.
Our method pioneers a decoupled training paradigm that strictly separates representation learning from noisy label supervision, fundamentally resolving the compounding confirmation bias inherent in existing noise-robust methods.
Specifically, we employ Momentum Contrast (MoCo) self-supervised training with RF-tailored augmentations to extract label-independent, hardware-intrinsic representations from complex-valued I/Q signals.
Building upon this feature space, a KNN-based noise filtering module isolates corrupted samples through neighborhood label consistency analysis, while an iterative rescue mechanism progressively recovers inadvertently discarded hard samples via joint evaluation of prediction confidence and prototype cosine similarity.
Comprehensive experiments on two real-world RF datasets demonstrate that SEI-SHIELD consistently achieves state-of-the-art identification accuracy across various noise rates, substantially outperforming existing regularization and sample-selection baselines.
The framework exhibits strong robustness to hyperparameter selection and maintains graceful degradation even under severe label corruption.
In future work, we plan to extend the framework to handle instance-dependent noise patterns, and explore lightweight contrastive architectures to reduce computational overhead.
\bibliographystyle{IEEEtran}
\bibliography{references}
\newpage

\section{Biography Section}
If you have an EPS/PDF photo (graphicx package needed), extra braces are
 needed around the contents of the optional argument to biography to prevent
 the LaTeX parser from getting confused when it sees the complicated
 $\backslash${\tt{includegraphics}} command within an optional argument. (You can create
 your own custom macro containing the $\backslash${\tt{includegraphics}} command to make things
 simpler here.)
 
\vspace{11pt}

\bf{If you include a photo:}\vspace{-33pt}
\begin{IEEEbiography}[{\includegraphics[width=1in,height=1.25in,clip,keepaspectratio]{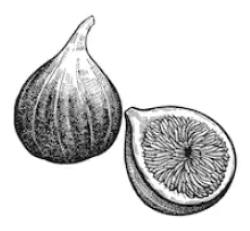}}]{Michael Shell}
Use $\backslash${\tt{begin\{IEEEbiography\}}} and then for the 1st argument use $\backslash${\tt{includegraphics}} to declare and link the author photo.
Use the author name as the 3rd argument followed by the biography text.
\end{IEEEbiography}

\vspace{11pt}

\bf{If you will not include a photo:}\vspace{-33pt}
\begin{IEEEbiographynophoto}{John Doe}
Use $\backslash${\tt{begin\{IEEEbiographynophoto\}}} and the author name as the argument followed by the biography text.
\end{IEEEbiographynophoto}

\vfill

\end{document}